\documentclass{easychair}

\usepackage{doc}
\usepackage{caption}
\usepackage{subcaption}

\newcommand{\fontsmall}{\fontsize{8pt}{9pt}\selectfont}

\usepackage{graphicx}
\usepackage{array}
\usepackage{longtable}

\usepackage{amsfonts}

\usepackage{colortbl}
\usepackage{tikz}
\usepackage{url}

\title{Benchmarking formalisms for dynamic structure system Modeling and Simulation}

\author{
Aya Attia
\and
   Clément Foucher
\and
   Luiz Fernando Lavado Villa
}

\institute{
 LAAS-CNRS, Université de Toulouse, UPS, Toulouse, France\\
  \email{\{aya.attia, clement.foucher, lflavado\}@laas.fr}
 }

\authorrunning{Attia, Foucher and Villa}

\titlerunning{Benchmarking formalisms for dynamic structure system M\&S}

\begin{document}

\maketitle

\begin{abstract}
Modeling and simulation of complex systems is key to explore systems dynamics. Many scientific approaches were developed to represent dynamic structure systems but most of these approaches are efficient for some kinds of systems and inefficient for others. Which approach can be adopted for different dynamic structure systems categories is a topic of interest for many researchers and until now has not been fully resolved. Therefore it is essential to explore the existing approaches, understand them, and identify gaps. To fulfil this goal, we identified criteria at stake for a smooth flow from model creation to its simulation for dynamic structure systems. Using these criteria, we benchmark the existing modeling formalisms focusing more on DEVS extensions, and use the results to identify approaches gaps and discuss them.
\end{abstract}


%
%

\section{Introduction}
Modeling and Simulation (M\&S) is a research field aiming at bringing a scientific approach to the creation of system models and the production of simulation results. M\&S offers a mathematical frame to represent systems as models and make these models evolve over time within a simulation environment. Models and simulators are powerful tools used in many scientific areas. Also, modeling is a critical concept in system design either for static systems or dynamic structure systems. This makes M\&S a cornerstone of many scientific and industrial fields.

Static systems are characterized by a fixed structure in which all possible events can be treated and fully described. Dynamic structure (or variable structure) systems are systems in which there is the possibility to add, change, remove parts of the system or modify coupling relations between them during simulation. In such systems, the conformity of the model regarding the real system is far more difficult to ensure than when dealing with static structure systems. This makes modeling dynamic structure systems among the most difficult challenges in M\&S, making them a specific topic of interest to researchers.

A formalism gives a frame to build models using a certain syntax and semantics as well as a set of rules allowed for their state changes during a simulation. A dynamic structure formalism needs to address the allowed ways of modifying the model structure and define a behavior to deal with these changes. The system changes should be made according to specified rules in the formalism. A great deal of research has been devoted to this goal. Many approaches developed for dynamic structure systems were reliable for some systems but inefficient for others.

In M\&S theory as defined by Zeigler~\cite{devs}, there are three main ways to describe models and simulators: using discrete time, using differential equations or using discrete events. From these, he proposes a mathematical descriptions for each one of these categories: Discrete Time System Specification (DTSS), Differential Equation System Specification (DESS) and Discrete Event System Specification (DEVS). These three are mathematically formal descriptions that are intended as being root of any other modeling and simulation technologies. Moreover, he proposes that DESS and DTSS can both be expressed using DEVS, which makes the later the equivalent of assembly language for simulators. DEVS is considered as the root in which all languages can be translated, but is often hidden behind more high-level languages.

DEVS offers a modular and hierarchical description of systems' dynamics using components. A system is represented as a set of connected components that can be atomic or coupled and react to external environment according to a set of rules. This approach ensures the reuse of models, thus minimizing rework and improve productivity. 

Initial research work on variable structure systems based on DEVS was DSDEVS made by Barros~\cite{dsde1998}. A special component called the executive network is responsible for changing structure and giving structure data of the systems. Many enhanced approaches were proposed to tackle the issues met with initial formalisms, such as DynDEVS~\cite{dynDEVS_uhrmacher_dynamic_2001}, its parallel version $\rho$DEVS~\cite{rouDEVS_uhrmacher_introducing_2006} or Cell-DEVS~\cite{wainer_advanced_2018}. These efforts aimed mainly at supporting system dynamics at all its levels and help system adaption with internal or external environments automatically.

Therefore, our contribution consists of analyzing existing formalisms and creating or updating a formalism that deals with issues encountered with the existing formalisms.

In this work, we focus our attention on deterministic dynamic structure systems literature, we try to extract and analyze formalism parameters used to handle dynamic structure systems. To do so, we need to extract main criteria that ensure the dynamic reconfiguration support.

The remaining of this article is organized as follows: a presentation of the basis of DEVS formalism is done in section~\ref{section_context}. The dynamic structure DEVS benchmarking is presented in section~\ref{benchmarking}, in this section we search for blocking reasons, determine main benchmarking criteria and present the benchmarking results. The benchmarking analysis are presented in section~\ref{benchmarking_analysis}. Our concluding remarks and future works are contained in section~\ref{conclusion}.

\section{Context of DEVS formalism}\label{section_context}

In this section we introduce the concepts used along this article. We first introduce the main concepts of DEVS formalism, then we represent some modeling formalisms including DEVS extensions and approaches outside of DEVS.

\subsection{DEVS}

\begin{figure}[tb]
    	\begin{centering}
    \includegraphics[width=8.5cm,keepaspectratio]{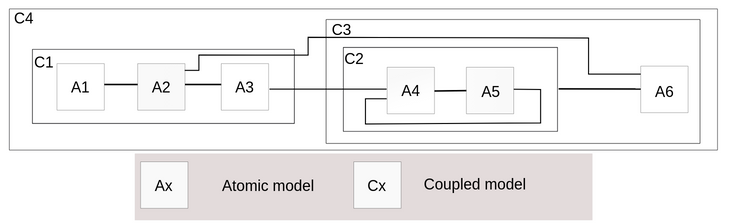}
    	\caption{DEVS atomic and coupled models}
    	\label{coupled_atomic_models}
    	\end{centering}
    \end{figure}

DEVS~\cite{zeigler_theory_2018} is a framework for modeling and simulation of complex systems using discrete events. It offers a framework based on mathematical concepts such as sets and systems theory to describe the structure and the behavior of a system. DEVS is able to represent complex real-world systems using a description of components hierarchically related and connected to each other following rules defined into the formalism. Components can either be atomic or coupled.
Atomic components present a behavior which depends on its internal state, that can evolve spontaneously with time passing, or in reaction to events on its inputs.
Coupled models combines atomic and/or coupled models to represent the relationships between different system components and propagate events between them. Figure~\ref{coupled_atomic_models} gives a simple overview of atomic and coupled models and its connections. 
When viewed from outside, coupled components act indistinctly from atomic, allowing to use coupled components the same way as atomics. This includes using them in other coupled, allowing for a hierarchical system description.

A CDEVS (Classic DEVS) atomic model~\cite{zeigler_theory_2018} is described by Equation~\ref{devs_atomic}:

\begin{equation}\label{devs_atomic}
DEVS = <X,Y,S,{\delta}_{ext},{\delta}_{int},{\lambda},ta>
\end{equation}    

\begin{minipage}{13cm}
\setlength{\parindent}{1pc} \fontsmall
\noindent Where:

$X = \{(p,v) \mid p \in InPorts, v \in X_{p}\}$ is the set of input ports and values,

$Y = \{(p,v) \mid p \in OutPorts, v \in Y_{p}\}$ is the set of output ports and values,

S is the set of sequential states,

\end{minipage}

\begin{minipage}{13cm}
\setlength{\parindent}{1pc} \fontsmall

$\delta_{ext} : Q \times X \to S$ is the external state transition function, with:

\begin{minipage}{11cm}
\setlength{\parindent}{1pc}
$Q = \{(s,e) \mid s \in S, 0 < e < ta(s) \}$ is the total state set,

e is the time elapsed since last transition
\end{minipage}

$\delta_{int} : S \to S$ is the internal state transition function,

$\lambda : S \to Y$ is the output function

$ta : S \to \mathbb{R}^{+}_{0} \cup \infty$ is the time advance function
\end{minipage}

In DEVS, a model is defined by a set of inputs X and outputs Y. Each element of these sets is defined by a pair of key and value and represent a potential event $v$ on a port $p$.
In the DEVS formalism, components state belong to the S set and can be changed using transition functions. When an input event occurs, $\delta_{ext}$ the external transition function is executed to change the component's state. If no external event occurs, the component will stay in its state until the period of time returned by time advance function ta(s) is reached, then it will change its state using the internal transition. When the state change, either as a result of internal or external event, the result of ta(s) will change accordingly. The component whose $ta$ is minimal amongst all components is said to be imminent. The output function $\lambda$ is executed when the state changed as a result of a component being imminent. After delivering the output, the imminent component list of the simulator is updated.

A DEVS coupled model~\cite{zeigler_theory_2018} is described by Equation~\ref{devs_coupled}:

\begin{equation}\label{devs_coupled}
N = <X,Y,D,\{M_{d} \mid d \in D\},EIC,EOC,IC,Select>
\end{equation}    

\begin{minipage}{13cm}
\setlength{\parindent}{1pc} \fontsmall
\noindent Where:

D is the set of the component names,

For each $d \in D$, $\{M_{d}\}$ is a DEVS model,
 
External input coupling connects external inputs to component inputs:

\begin{minipage}{11cm}
\setlength{\parindent}{1pc}
$EIC \subseteq \{((N,ip_{N}),(d,ip_{d}) \mid ip_{N} \in IPorts_{N}, d \in D, ip_{d} \in IPorts_{d} \}$
\end{minipage}

External output coupling connects component outputs to external outputs:

\begin{minipage}{11cm}
\setlength{\parindent}{1pc}
$EOC \subseteq \{((d,op_{d}),(N,op_{N}) \mid op_{N} \in OPorts_{N}, d \in D, op_{d} \in OPorts_{d} \}$
\end{minipage}

Internal coupling connects component outputs to component inputs:

\begin{minipage}{11cm}
\setlength{\parindent}{1pc}
$IC \subseteq \{((a,op_{a}),(b,ip_{b}) \mid  a,b \in D, op_{a} \in OPorts_{a}, ip_{b} \in IPorts_{b} \}$
\end{minipage}

$Select : 2^D \to D$ the tie-breaking function
\end{minipage}


Coupled models dictate the system composition by describing how components will interact during simulation. In a coupled model a set of the network inputs X and outputs Y are defined. The D set contains the names of the components contained into this model. A coupled model connects the components to each other thanks to connections defined into EOC, EIC, and IC sets. The select function is used for tie-breaking if there is more than one imminent component at a given time.

CDEVS does not support multiple simultaneous inputs, so it was essential to develop a parallel approach. PDEVS was developed to tackle this problem and provide more flexible alternative. PDEVS~\cite{pdevs} is a parallel version of DEVS that replaces the select function with another mechanism to handle parallel execution. It allows many components to evolve simultaneously and execute concurrent events. $\delta_{con}$ is a function added to PDEVS, this function deals with confluent situation while simulating a system.

A PDEVS atomic model is described as~\cite{zeigler_theory_2018}:

\begin{equation}
DEVS = <X^+_{M},Y^+_{M},S,{\delta}_{ext},{\delta}_{int},{\delta}_{con},{\lambda},ta>
\end{equation}    

\begin{minipage}{13cm}
\setlength{\parindent}{1pc} \fontsmall
\noindent Where:
    
$\delta_{con} : Q \times X^+_{M} \to S$ is the confluent transition function

\end{minipage}

In the PDEVS formalism, $X^+_{M}$ and $Y^+_{M}$ represent multiple inputs and outputs of the model, the $^+$ indicates that multiple events can be present instead of a single event at a time with CDEVS. PDEVS allows to receive multiple inputs and treat them simultaneously. Inputs received when a component is imminent (i.e $e=ta(s)$) result in collisions. $\delta_{con}$ is there to handle collisions, this function processes collision behavior and handles them by defining instructions that should be applied in these situations. Most of the time, $\delta_{con}$ calls $\delta_{int}$ and/or $\delta_{ext}$ in a specific order, but can also be more complex. Thanks to this function, models are fully controlled in collision. A PDEVS coupled model is the same as in DEVS except for the select function which is removed~\cite{zeigler_theory_2018}.

\subsection{DEVS Simulation}

The previous formalism gives a formal representation of the system that is based on mathematical sets and functions. This representation should be simulated to show system behavior.

\begin{figure}[!ht]
  \centering
  \begin{subfigure}[b]{0.45\textwidth}
    \centering
    \includegraphics[height=4.2cm]{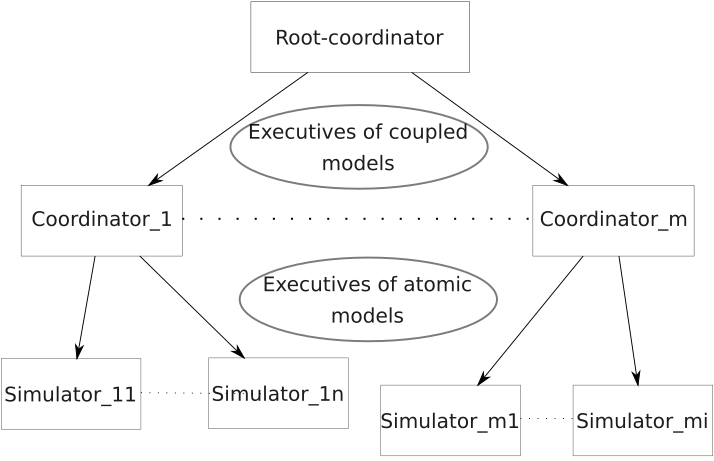}
    \caption{Simulation hierarchy}
    \label{simulation}%
   \end{subfigure}
   \hfill
   \begin{subfigure}[b]{0.45\textwidth}
    \centering
    \includegraphics[height=4.2cm]{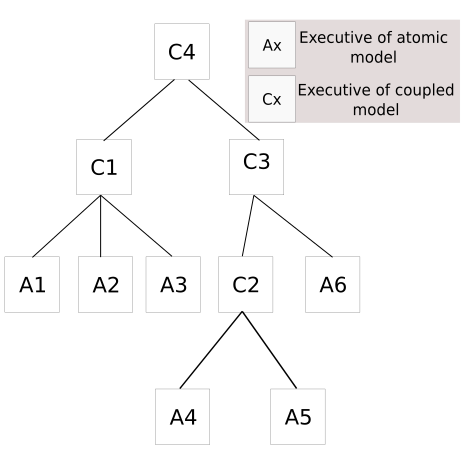}
    \caption{Concrete simulation example}
    \label{concrete example}%
  \end{subfigure}
  \caption{Simulation abstraction and concrete example}
  \label{sim_hierarchy}
\end{figure}

    In the abstract simulator concept~\cite{chow_parallel_1996}, the simulation of atomic and coupled components is carried out by processors called respectively simulator and coordinator. Additionally, a root coordinator schedules the system simulation. These processors support the representation of model behavior. Figure~\ref{simulation} presents the general hierarchy of a simulation and Figure~\ref{concrete example} shows a concrete example for the DEVS model described into Figure~\ref{coupled_atomic_models}.

    The simulation begins by an order of initialization delivered by the root coordinator to its subordinates. Later, its role is to activate imminent components. The root coordinator then advances the simulation time to match the $ta$ of the imminent(s) component(s)~\cite{zeigler_theory_2018}. An activated imminent component will produce an output event just before changing its state. The coordinator is in charge of gathering output events from its inner components, and deliver them as input event to connected components. In addition, the coordinator updates a list of imminent components, the time of last event $t_{last}$ and computes time of next event $t_{next}$. Furthermore the simulator supports other tasks: it computes variables like the time of last event $t_{last}$, time of next event $t_{next}$, elapsed time e and update model’s state. 
    
\subsection{Modeling formalisms}

In this section we give a brief description of modeling formalisms extended from DEVS and other approaches.

\subsubsection{DEVS extensions}

    Many extensions for DEVS were proposed to enhance existing approaches. As an example, DEVS is limited to modeling deterministic systems. To overcome this, an extension was developed to support stochastic models is the STochastic Discrete Event System Specification (STDEVS)~\cite{kofman_stdevs_2006, castro_formal_2010}. 
    This approach performs stochastic space calculations to determine next states. STDEVS was powerful for many systems such as Wind farms~\cite{perez_stochastic_2010} and load-balancing system~\cite{castro_formal_2010}. In this article, we will focus on dynamic structure extensions. 
    
    
    Despite the ability of existing static approaches to represent system execution, it is difficult to deal with dynamic structure changes of the system. This problem makes proposed formalisms inefficient for some cases. Therefore, many research works focus on creating efficient approaches for dynamic structure systems.
    
    The first formalism proposed for modeling dynamic structure systems is Variable DEVS~\cite{barros_variable_1994}. This formalism is composed from atomic DEVS models and coupled models which contain a couple named $\chi = (M, C)$ which itself contains models ($M$) and their connections ($C$). But in this approach, structure decision maker is not defined therefore it isn't considered efficient.  
    
    To represent and simulate systems that undergo structural changes, Barros then proposed the DSDEVS formalism~\cite{dsdevs_barros_dynamic_1995}. This formalism is composed of an atomic and network executive components (a modified atomic model that specifies structural changes). The network executive has a global overview of all possible architectures of the system, it stores all possible states of structural changes and their corresponding component sets in each structural state~\cite{shang_simulation_2006}. A weakness of the DSDEVS is that it can not support multiple simultaneous events, so a parallel DSDEVS named DSDE~\cite{dsde1998} was proposed. Thanks to parallelism, the system can deal with multiple inputs and is able to provide outputs simultaneously.
    
    DynDEVS~\cite{dynDEVS_uhrmacher_dynamic_2001} and $\rho$DEVS~\cite{rouDEVS_uhrmacher_introducing_2006} are well used in the fields of biology. DynDEVS is an extension of the DEVS formalism adding model transition functions that allow structural changes by both atomic and coupled components (as opposed to DSDEVS where structure changes are only possible at the coupled level). DynDEVS cannot support multiple simultaneous inputs therefore $\rho$DEVS was proposed which adds dynamic ports to formalism and allows parallelism. These formalisms are used in different fields like biology~\cite{rouDEVS_uhrmacher_introducing_2006} or Agents modeling using JAMES~\cite{uhrmacher_role_2002}.
    
    Cell-DEVS~\cite{wainer_advanced_2018} is among well used formalisms. It is an extension of Cellular Automata and DEVS that are combined together to take advantage of both. This formalism has efficient strategy to represent system dynamics, it divides systems into cells that represent the atomic models and grid that contains cells related to each other which represent coupled models. 
    
    Cell-DEVS is commonly used: among its applications are modeling of residential neighborhood, commercial neighborhood~\cite{wainer_application_2001}, evacuation and crowd modeling, 
    bacteria~\cite{wainer_survey_2010}, 
    surface tension~\cite{wainer_advanced_2018}, 
    pedestrian movement in a building or computer malware~\cite{wainer_advanced_2018}.
    
    Another approach, hybrid-DEVS~\cite{pawletta_devs-based_2006}, suggests that discrete event system simulation cannot effectively simulate systems, so it combines discrete and continuous phenomena. The main goal of this formalism is to represent the dynamic structure in discrete event/continuous systems. This formalism was notably used in semiconductor manufacturing supply chain systems~\cite{huang_flexible_2006} or elevator controller~\cite{saadawi_hybrid_2013}.
    
    Sometimes it is good to combine some existing formalisms to derive the full features. A formalism named Extended Dynamic Structure DEVS (EDEVS)~\cite{hagendorf_extended_devs_2009} was proposed on the basis of existing DEVS formalisms. In this formalism, the missing elements in DEVS found in other formalisms extended from DEVS were exploited like PDEVS, DSDEVS and CDEVS.

\subsection{Formalisms outside of DEVS}
      
    DEVS is not the only tool used for Modeling and Simulation. In this work, we include some approaches outside of DEVS for comparison methodology. For example SysML~\cite{hause_omg_2007, graves_using_2011, gauthier_sysml_2015} was also used for many fields. SysML is a  modeling language that extends UML which provides formal semantics and support various levels of complex systems development. This M\&S tool was used for embedded mechatronic system domain~\cite{ambert_applying_2013}, also in Intelligence, Surveillance and Reconnaissance design~\cite{graves_using_2011}.
    
    Another approach named Variable Structure System Specification (VSSS)~\cite{lee_methodology_1997} was proposed to deal with variable structure systems. This approach had a similar mindset of DEVS, but it differs in how the specification of system structure is made. Two characteristic functions are defined in the compositeVSSS named \textit{f} and \textit{g} which represent state transition and mapping functions which map models under given state to specify system structure.

\section{Dynamic structure formalisms benchmarking}\label{benchmarking}

Each of the approaches presented in previous section was efficient for some dynamic structure systems and inefficient for others. Indeed, an efficient dynamic approach should cover all possible dynamic structure system problems. As result, it is essential to understand these approaches, analyze it to find gaps and extract main criteria. In this section, we present a benchmarking of dynamic structure DEVS systems.

    \subsection{Benchmarking motivation}\label{benchmarking_motivation}
    
    In dynamic structure systems, on one hand, it is important to understand the dynamical nature of the components and interactions. On another hand, it is essential to have knowledge about the development of the underlying network topology and the hierarchy of component dynamics. Variable structure systems are characterized by dynamic individual and collective behaviors that should be expressed using the formalism.
    
    
    One problem is that the rules in the existing modeling approaches restrict the evolution of component dynamics. In some cases, the approach rules does not support some kinds of events that aren't described into the formalism. So it becomes impossible to update the influenced components and support the dynamic structure changes.
    
    As inputs to the dynamic structure systems are varied and sometimes it is hard to identify them, connection strengths change and learning takes place~\cite{gorochowski_evolving_2012} but this is not possible with almost all existing approaches because these approaches contain predefined rules and do not have the ability to deal with blocking situations except if concrete behavior is added to these approaches which is not the case currently.
    
    In addition, one type of structure representation optimized for one field of expertise does not necessarily holds for other fields. Thus, a generalized approach is needed to handle this variety when describing systems whose structure changes over time. A generalized approach should offer a common description for many types of complex dynamic structure systems in applied science and engineering.
    
    Furthermore, DEVS presents a well defined formalism based on mathematical rules which are strictly correct and clear. With a formalism purely based on mathematical rules, the correctness of the transformation from a mathematical model to an executable model can present a challenge. Errors can then emerge due to the programmer, to the simulator itself, or to the impossibility to exactly match the formal model to an executable model. For example in some cases, additional variables will be introduced for temporary computations. Or some specific details from the formalism will be lacked by the simulator. It can even be that the language used by the simulator will allow for shortcuts that break the formalism such as changing the model state outside of delta function in DEVS.
    
    Problems encountered include the transformation from abstract models (mathematical/ formal models) to executable models, this transformation may not comply with rules in the abstract model. Indeed we find that variables definition in abstract model is widely different from its definition in executable model. In abstract model, variables are introduced as sets of allowed values, without a name. But in the executable model, each variable should be fully defined using a name and, depending on language, a type. For instance, in C++ programming language variables have an immutable type. In python, variables have a name, but the type can change dynamically and thus not match the definition set. This creates gaps between abstract and executable models. So, dynamic structure systems need a formalism that define the set of components or agents, the interactions that take place between them and, the main ingredients of dynamic system, capture and describe the types of changes that can occur.
    
    Therefore, to represent dynamic structure systems, a fully detailed approach will be necessary.
    As a step toward this goal, a benchmarking of dynamic structure formalisms is made in the next section in order to analyze the existing approaches and extract important features that manage the dynamic behavior.

    \subsection{Benchmarking}\label{subsection_benchmarking}
    
    To do our benchmarking, we need to identify benchmarking criteria and criteria weights used in dynamic structure systems benchmarking.

    \subsubsection{Benchmarking criteria}\label{benchmarking_criteria}
    
    Several possible criteria were suggested to describe system dynamics during all phases of its development and interactions with external events. In this section, we define our main criteria for dynamic structure systems benchmarking, namely, distribution of decision-making authority, structure information, ports and port modification.  
        
    Decision-making authority takes an important part in describing system dynamics. The centralization strategy was considered in many existing approaches as an important criterion that is responsible to take decisions making authority. Some formalisms give this authority to all components and in others it is restricted to special components, this affect system interactions with external world. 
         
    Structure information is important to provide a detailed representation of system composition, interactions, dynamics. To better represent the structure changes, it is essential to provide detailed description of different elements in the system such as components information like their names, types, rules to create, update or remove components. Furthermore the description of components behavior using transition functions give an overview of different system interactions. 
    
    Port declaration is essential to specify if the approach defines ports into its formalism. This criteria help developers to allocate ports, assign inputs, understand the mechanism of system flows. 
    Furthermore the port specification gives an overview of system development flows. Specifying different components interface facilitate the manipulation of the system with external and internal environment and help to manage it correctly.
    
    Finally, ports modification helps to show adaption of the system with structural changes. Ports with variable property support structure updates and deals with coupling changes. The ports modification allows to change dynamically models interface which improve the dynamic adaption during system structural changes. In addition, this gives high level of compatibility in the description of abstract model in relation to the modelization of real time systems. Indeed, real time systems endure multiple changes according to their current state, received inputs at real time that can be deleted, added or updated.

    These criteria are used to benchmark the existing variable structure approaches in which we attribute a value between 0 and 1 for each criterion. This value represents the conformity of the approach with the mentioned criterion as follows:
    \begin{itemize}
     \setlength\itemsep{-0.7em}
        \item 0 means no conformity
        \item Values in $\left]0;1\right[$ means that the approach tackles some details of the criteria but not all
        \item Value 1 means that the approach tackles all details of the criteria
    \end{itemize}
    The value obtained for each criterion is then multiplied by the criterion weight.

    \subsubsection{Benchmarking weights}\label{benchmarking_weight}
    
    Each criterion used in the benchmarking should have a weight reflecting its importance according to our goals. Using this benchmarking, we specified which approach is the most suitable according to dynamic structure systems criteria. To do so, we assign a score to each benchmark criteria as mentioned below: distribution of Decision-Making authority: 3 points, structure information: 3 points, existence of ports: 4 points, ports modification: 1 point.
    
    Indeed the existence of ports criterion has most important weight because it specifies the different flows during the simulation of the system. Then distribution of Decision-Making authority strategy and structure information criteria have also an important weight because these criteria give an idea about system's execution mechanism during all levels. The criterion of ports modification has low weight because this information is detailed in system's structure information and modeling and simulating system can be well defined without this criterion.

    \subsubsection{Benchmarking Results}\label{benchmarking_results}
\begin{figure}[!ht]
    	\begin{centering}
    \includegraphics[height=5.5cm,keepaspectratio]{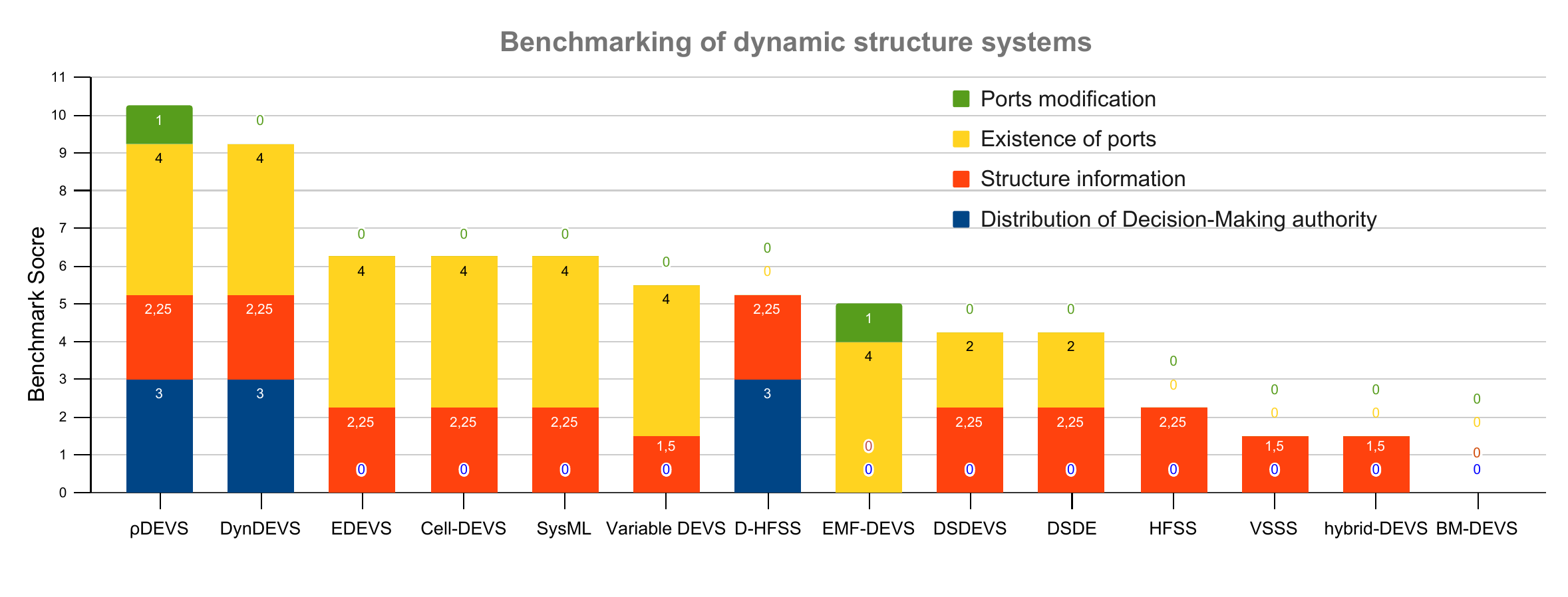}
    	\caption{Benchmarking of dynamic structure systems}
    	\label{benchmark_table}
    	\end{centering}
    \end{figure}

    After specifying the score of each criterion, we compute the score of the presented approaches. Figure~\ref{benchmark_table} presents results of our benchmarking.

    After computing scores, we find that Variable-DEVS, DynDEVS, $\rho$DEVS, EDEVS, Cell-DEVS, DynDEVS and SysML are the most interesting approaches thanks to their scores which range between 5.5 and 10.25. Based on these scores, we select $\rho$DEVS and Cell-DEVS as most helpful formalisms. They can be used as a base to create another efficient formalism for dynamic structure systems. Note that, while DynDEVS has a higher score than Cell DEVS, we didn't select it as its evolution, $\rho$DEVS was already selected. Also EDEVS, Cell-DEVS and SysML have the same score but we select Cell-DEVS thanks to its description and its details found into the formalism, it is more used than EDEVS and we aim to draw upon DEVS.

\section{Benchmarking analysis}\label{benchmarking_analysis}

In this section we analyze the benchmarking results which help us to identify Modelization problems.

\subsection{Identified modelization problems}\label{modelization_pb}

The previous section presented a benchmarking of a set of formalisms for dynamic structure systems. Although these formalisms deal with some variable structure systems, many types of inconsistency and conflict are possible. In this section, we show the main reasons that generate theoretical issues raising the need for a reconsideration of the existing dynamic structure DEVS formalisms. Issues come from formalism gaps such as the centralization strategy.

Centralized systems give structure changes authority to coupled models only like with Variable DEVS~\cite{barros_variable_1994}, DSDEVS~\cite{dsdevs_barros_dynamic_1995}, DSDE~\cite{dsde1998} in which structure changes authority is limited only to the executive model (the executive model presents a hybrid form between atomic and coupled models and is responsible for structure changes and storing it). Centralization allows to have a single reference and avoids decision conflicts but it presents a single point of system bottleneck. To avoid centralization risks, many approaches trend to distribute this authority to multiple components. Atomic and coupled models can collaborate and make structure decisions like in DynDEVS approach~\cite{dynDEVS_uhrmacher_dynamic_2001}, $\rho$DEVS~\cite{rouDEVS_uhrmacher_introducing_2006} and D-HFSS~\cite{barros_representation_2014} in which atomic and coupled models have the right to change system's structure by adding, updating and deleting models. This strategy offers several benefits, including improved reliability, scalability and performance. 

Among criteria that is not well defined in many approaches is the system's structure data. Indeed, it is essential to define some details about system structure like the allowed components and their relationships. System structural change rules should be detailed such as component addition, removal, replacement, and network reorganization. Initialization takes part of most important structure change details which should be considered from the beginning. One of the reasons that explain the importance of it is because of the need of initialization when restarting a simulation run after it was interrupted. The state has to be re-initialized to the last know state, from which simulation can subsequently resume as if it was never interrupted~\cite{van_tendeloo_extending_2018}. DynDEVS~\cite{dynDEVS_uhrmacher_dynamic_2001} is considered among the most clear approaches, in this formalism initial state was defined in which we find the detailed minimal composition of the network thanks to $m_{init}$ set. But having a static definition of the initial state cannot be sufficient for all cases. Indeed, initial state can be changed depending on the current state of other components and at the same time a new component is added, its composition and time i.e the initialization in booting system differ from initialization after executing some parts of the system. In reality the initial state of added components at execution time differ from the component's initial state during booting the system. This problem was ignored by the existing approaches which explain the complexity of the implementation of these approaches in some cases. In addition, the initialization is highly required when the simulation is interrupted due to some system condition, the model is altered, re-initialized to a consistent state, and the simulation resumed~\cite{van_tendeloo_extending_2018}. Therefore, the initial state should be computed dynamically depending on current system values.

The system's structure data is not limited to initial state but there are many important details that should be detailed into the formalism. It is essential to know how the structure changes are made and the description of its different possible use cases while defining transition functions and the state set description. The DynDEVS formalism~\cite{dynDEVS_uhrmacher_dynamic_2001} describes how structure changes are made: to change structure model transition function is executed but this function does not describe all changes that can be made.

Cell-DEVS~\cite{wainer_advanced_2018} presents clear descriptions for atomic and coupled models: atomic models are presented as cells in which each cell has its interface that defines the number of cells and describes the connections between cells and its neighbours. The same concept is used for coupled models and adds extra elements like $C$ set which represents the cell space also the $B$ set which contains details about borders. This information gives a global overview about the system dynamics which help to understand how system can be simulated and fix problems but these are insufficient because some other important details (initial state, ports types, components descriptions) should be specified into the formalism. 

The description of system's structure may lead to structural and behavioral problems: inaccuracy in the definition of states, how states are computed after each transition, how states are exactly introduced like its types, values that can be present, system's composition, relations between components. Also, the specification of domain of variables is of real significance for system dynamics guidance but it is ignored by many approaches.

Furthermore, in the definition of ports which is important to specify the type of different flows and have organized knowledge about system flows. Indeed, we found that the most of formalisms don't define ports and its types which lead to problems of inconsistency in some simulation cases. Ports definition is important to give an overview about possible flows while receiving inputs or returning outputs. Many approaches ignored ports concept like DSDEVS~\cite{dsdevs_barros_dynamic_1995}, DSDE~\cite{dsde1998}, VSSS~\cite{lee_methodology_1997} and other approaches, which explain the faced conflicts during some simulations. Also, in some cases we found inaccuracy in the definition of domain of variables.

In addition, it is notable that the transformation from abstract model to executable model may be inconsistent because developers face many difficulties in translating abstract model to executable code. The majority of approaches doesn't specify the variable type for inputs or outputs which lead to difference between what exist in abstraction and in execution. Also, in some cases developers are forced to add some details like internal variables, add functions and other additional parts or removing some elements like in ADEVS~\cite{adevs} simulator in which the $I$ set that represents the influencers is not defined and there is an added function named route function which guide Inputs/Outputs of each component. These changes generate inconsistency between abstraction and execution. ADEVS~\cite{adevs} is not the only approach that take freedom during the transformation from abstract model to executable model, in CD++~\cite{noauthor_cadmium_2023} \cite{cd++_2022} some functions are ignored and others are added. The same problem with PythonPDEVS simulator~\cite{l_installation_2019} in which ModelTransition function is used instead of the $C_\chi$ component. 

Issues come from formalism gaps such as the specification of structure decision making authority, poorly defined structure data, ports definition and inconsistent transformation from abstract model to executable model.

\subsection{Discussion}\label{Discussion}

Considering existing formalisms, we find many gaps for critical points: because it is easier to implement, centralization concept was preferred by most of formalisms like Variable DEVS~\cite{barros_variable_1994}, DSDEVS~\cite{dsdevs_barros_dynamic_1995}, DSDE~\cite{dsde1998} in which structure changes authority is limited only to the executive model, also in hybrid-DEVS~\cite{pawletta_devs-based_2006}, EDEVS~\cite{hagendorf_extended_devs_2009}. A fewer number of approaches adopt opposite strategy by sharing this authority like DynDEVS~\cite{dynDEVS_uhrmacher_dynamic_2001}, its parallel version $\rho$DEVS~\cite{rouDEVS_uhrmacher_introducing_2006} and D-HFSS~\cite{barros_representation_2014}. The trend to centralization for many approaches explain blocking reasons in some situations, the only authority that guide simulator during its execution is blocked so the whole is blocked. Therefore, few number of approaches process this issue and try to find solution by sharing this authority between its components like DynDEVS~\cite{dynDEVS_uhrmacher_dynamic_2001} in which atomic and coupled models share this right and are able to take structure decisions that decrease the problem of single point of failure in simulator and decrease the risk of error.

Considering system's structure data, almost all approaches give information about structure which offers an overview about system simulation. This information was presented by most of approaches like DSDEVS~\cite{dsdevs_barros_dynamic_1995}, DSDE~\cite{dsde1998}, hybrid-DEVS~\cite{pawletta_devs-based_2006}, EDEVS~\cite{hagendorf_extended_devs_2009}, DynDEVS~\cite{dynDEVS_uhrmacher_dynamic_2001}, Cell-DEVS~\cite{wainer_advanced_2018}, HFSS~\cite{barros_representation_2014} and ignored by a very few number of approaches like BM-DEVS~\cite{bmdevs_cho_simulation_2020}. This information is very important to understand system dynamics. Given insufficient system's structure data leads to incoherence in the transformation from abstract to executable model. To effectively simulate the system additional elements are added during the flow from abstract to executable model for example in Cell-DEVS~\cite{wainer_advanced_2018} simulator there are some additional functions like setLocalTransition or inverseNeighborhood in simulator code. In addition, some information in the formalism are considered as extra data 
in simulation and are ignored. For example, in the same simulator mentioned above, the select function and the borders set do not exist in the simulator code although borders set is considered important thanks to its ability to provide an overview of influencers and influencees and which differ effectively while computing transition functions. 
Among identified inconsistencies the description of DSDEVS~\cite{dsdevs_barros_dynamic_1995} formalism in which some details aren't mentioned into abstract model but are found into executable model like the input and output sets description of these two kinds of models. Indeed the input and output of abstract models are described that are sets of values without specifying ports but into executable model we found the declaration of ports. 
Also in the same example, we found some functions into simulator that does not described into the formalism like addModel, removeModel, replace, link, unlink, find, clear and replace. 

In most cases, the transformation from abstract model to executable model is done by a manual procedure. The formalism itself is transformed into executable code as simulators. The description of transition functions, output function are described into simulator but in some cases the simulators need to introduce additional variables or functions that help the simulator to achieve desired goal. This produces executable models that do not respect all rules set into the abstract model which reduces confidence in this transformation.
Some other approaches try to cover this issue by suggesting translators like VSSS~\cite{lee_methodology_1997} approach in which a translator is proposed in order to ensure vertical transformation from abstract model to executable model. This transformation ensures similarity in abstract and executable model.

The definition of ports is important knowledge to have about system flows. This concept was treated by some approaches like Variable DEVS~\cite{barros_variable_1994}, DynDEVS~\cite{dynDEVS_uhrmacher_dynamic_2001}, its parallel version $\rho$DEVS~\cite{rouDEVS_uhrmacher_introducing_2006}, EMF-DEVS~\cite{sarjoughian_emf-devs_nodate}, Cell-DEVS~\cite{wainer_advanced_2018} and SysML~\cite{gauthier_sysml_2015} and ignored by others like DSDEVS~\cite{dsdevs_barros_dynamic_1995}, DSDE~\cite{dsde1998}, VSSS~\cite{lee_methodology_1997}, HFSS~\cite{barros_representation_2014} and D-HFSS~\cite{barros_representation_2014}. Indeed these approaches don't specify if there are ports specification or not.

Ports types is only supported by Cell-DEVS~\cite{wainer_advanced_2018} but the other approaches do not take attention about the type of variables received into its ports. Few number of approaches take care of the definition of ports at root like in Variable DEVS~\cite{barros_variable_1994}
, Cell-DEVS~\cite{wainer_advanced_2018} and SysML~\cite{gauthier_sysml_2015}.

Systems that change dynamically need to have dynamic equipment like ports. Indeed, it is highly required to have ports that can be changed dynamically i.e ports can be added when the system dynamics need additional ports to continue its execution, can be deleted also when the existing ports are not useful, in some other cases ports definitions need to be changed like updating ports value type. This concept was treated only by $\rho$DEVS~\cite{rouDEVS_uhrmacher_introducing_2006} and EMF-DEVS~\cite{sarjoughian_emf-devs_nodate}. The ports modification is an additional part to DynDEVS to create $\rho$DEVS and which make $\rho$DEVS more useful than DynDEVS. The other approaches do not take attention to this concept, this explain the trend to apply $\rho$DEVS for many systems according to other existing approaches. So it is useful that the approach suggest port modification i.e the approach suggest if there are ports added, removed or updated dynamically.

Using analytical results in Figure~\ref{benchmark_table}, we can observe that no single formalism covers the whole criteria for modeling dynamic structure systems. Although none of the existing formalisms implements all the required features, a combination of different formalisms can be operated to extract useful elements of the modeling process and create an efficient formalism.
In search for a proper formalism, perhaps the most important aspect to consider is the balance between simplicity and expressiveness~\cite{machado_modeling_2011}. To fulfil this goal, the state set, formalism functions, all useful details of the system should be expressive and simple. This helps to create efficient modeling approach. Efficient modeling approach use a distributed strategy that avoid blocking risk, also it is important that all structure details are mentioned into the formalism. Giving all modelization details gives an overview about system dynamics and helps developers to create the executable model. We find also that it is essential to specify ports and describe its characteristics. Providing good formalism enables to have strong system description and ensures efficient simulation.

\section{Conclusion and future work}\label{conclusion}

The main objective of our research is to understand the strategy used in the existing approaches for dynamic structure systems, identify gaps and provide an adaptive formalism for dynamic structure systems. To do so we extracted the main criteria for the development of dynamic structure systems used in literature then we used these criteria to present dynamic structure system formalism benchmarking. The benchmarking was helpful to understand the strategy used by the existing approaches and analyze it. 

By analyzing the benchmarking results we identified that among reasons for dynamic structure systems inefficiency is forgetting details in the description of the formalism, ports existence and its description. For future work, we intend to provide a proof of correctness in the description of each variable in the formalism by providing the maximum of
important details of system description as possible. This helps to avoid this kind of problems. The consistency between these two kinds of models should be included among the formalism rules. To do so, the formalism should be fully defined up to the executable model and its execution, and strict rules should be fixed in order to ensure vertical transformation that will be a part of our future works.

Using these analysis, we are able to select efficient strategy to develop or update an efficient formalism for dynamic structure systems. The research works to have efficient formalism for dynamic structure systems will continue because so far the existing formalisms still has a lot of gaps and this type of formalisms is highly needed in real world. Furthermore our research works will try to provide tools that facilitate the M\&S process and promote the adoption of formal M\&S approaches by a wider community of practitioners and researchers.

\label{sect:bib}
\bibliographystyle{plain}
\bibliography{easychair}

\end{document}